\documentclass[prl,aps,twocolumn,bibnotes,footinbib]{revtex4-2}
\usepackage{amsmath,amssymb}
\usepackage[colorlinks,bookmarks=false,citecolor=magenta,linkcolor=magenta,urlcolor=magenta]{hyperref}
\usepackage{tikz}
\usepackage{mathtools}
\usepackage{bm}
\usepackage{ragged2e}
\usepackage{graphicx}
\usepackage[symbol]{footmisc}

\begin{document}
	\title{Dissipative preparation of Laughlin-like states}
	\author{Tong Liu}
	\email{tliu.phys@gmail.com}
	\affiliation{Chalmers Next Labs, 41296 Gothenburg, Sweden}

	\begin{abstract}
	Fractional quantum Hall (FQH) states are a central paradigm of strongly correlated quantum matter and a key platform for topological quantum computation.
	Here, we propose a purely dissipative protocol based on local loss and pump channels for preparing Laughlin-like states at filling $1/3$, with a possible extension to other $1/M$ filling states.
	We show that the Laughlin-like state is the exact unique steady state of the Lindbladian under open boundary conditions.
	Finite-size analysis of the Lindbladian gap suggests efficient dissipative preparation over the system sizes and parameter regime considered.
	We further demonstrate adiabatic pumping of a Laughlin-like state through slow modulation of the pump channels during the evolution.
	Our work opens a feasible route to preparing and manipulating FQH states on near-term quantum simulators.
	\end{abstract}

	\maketitle

	Confined to a two-dimensional surface threaded by a strong magnetic field, electrons enter a regime in which interactions dominate over the kinetic energy, forming an incompressible liquid known as fractional quantum Hall (FQH) state~\cite{PhysRevLett.50.1395,RevModPhys.71.S298}.
	Fractionally charged quasiparticles obeying anyonic statistics are the building blocks of topological quantum computation~\cite{PhysRevLett.53.722,RevModPhys.80.1083}.
	Using adiabatic evolution and quantum circuits, FQH states have been realized on programmable quantum platforms with particle-resolved access to the microscopic structure of many-body wavefunctions~\cite{Clark2020,PhysRevLett.129.056801,Lonard2023,Wang2024,Shen2026}.
	However, unavoidable environmental coupling limits the achievable state fidelities by constraining the available evolution time in adiabatic preparation and the maximum length of quantum circuits, posing a challenge for scaling such methods to larger systems.

	Engineered dissipation provides a complementary approach to quantum-state engineering by tailoring the relaxation dynamics of open quantum systems~\cite{PhysRevA.78.042307,Diehl2008,Verstraete2009}.
	In contrast to coherent protocols based on controlled unitary evolution, dissipative preparation can make the target state a dynamical attractor, enabling autonomous convergence from a broad class of initial states and a degree of intrinsic stabilization against perturbations.
	It has been exploited to prepare states ranging from few-body entangled states~\cite{Shankar2013,Lin2013,Barreiro2011} to correlated many-body states~\cite{Ma2019,Mi2024}.
	Recent theoretical proposals have explored the dissipative preparation of bosonic FQH states in lattice models with synthetic magnetic fields~\cite{PhysRevLett.108.206809,PhysRevX.4.031039,PhysRevA.96.053808,PhysRevResearch.3.043119,Kurilovich2022,2026arXiv260518377S}, typically through an interplay between coherent dynamics and dissipation.

	Here, we develop a purely dissipative scheme for preparing Laughlin-like states.
	We show that local two-particle loss and conditional pumping processes select the Laughlin-like state as the exact unique steady state of the open system under open boundary conditions.
	We focus on the states at filling fraction $\nu = 1/3$, while the protocol can be generalized to other $\nu=1/M$ states, including bosonic $\nu=1/2$ states.
	By combining full-space exact diagonalization with a branch-resolved analysis, we identify the decay modes that control the Lindbladian gap across the explored parameter range.
	The finite-size analysis suggests efficient dissipative preparation over the system sizes and parameter regime studied.
	We also demonstrate continuous pumping of a Laughlin-like state by slowly modulating the pumping processes.
	Finally, we outline a possible implementation with neutral-atom platforms, providing a concrete path toward realizing dissipative preparation of FQH states in experiments.

	\begin{figure}[t]
		\centering
		\includegraphics[width=\linewidth]{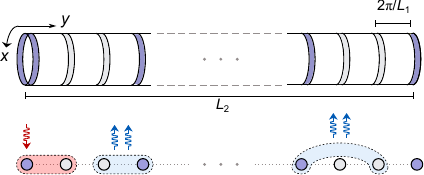}
		\caption{\quad A schematic of dissipative preparation of Laughlin-like states on a cylinder. 
		After projection onto the lowest Landau levels, these states can be prepared by local loss and pump channels.}
		\label{fig:cartoon}
	\end{figure}

	\emph{Model.}
	We consider $N_e$ spin-polarized fermionic particles restricted to the lowest Landau levels (LLL) on a cylinder of circumference $L_1$ in the $x$-direction and length $L_2$ in the $y$-direction~\cite{PhysRevB.31.2529,PhysRevB.50.17199}.
	For simplicity, we choose the Landau gauge, $\bm A = -By\hat x$, and set the magnetic length $\ell_B = \sqrt{\hbar/|q|B} = 1$ where $q$ is the particle charge.
	As depicted in Fig.~\ref{fig:cartoon}, the LLL orbitals $|n\rangle$, labeled by the quantized momentum along the $x$-direction, are localized in the $y$-direction at $y = n\kappa$, with $\kappa = 2\pi/L_1$ and $n = 0, 1, \dots, N_\phi-1$, where $N_\phi = L_1L_2/2\pi$ is the number of flux quanta piercing the cylinder and hence the total number of available orbitals. 
	To be consistent with the topological structure of the cylinder, the filling faction is defined as $\nu = (N_e - 1)/(N_\phi - 1)$.
	Within the LLL subspace, the interaction Hamiltonian can be written in terms of Haldane pseudopotentials as $H = \mathcal \sum_{i<j}\sum_{m} \mathcal U^m \mathcal P_{ij}^m$ where $\mathcal P_{ij}^m$ denotes the projector onto the two-particle channel associated with relative momentum $m$, and $\mathcal U_m$ is the corresponding pseudopotential coefficient~\cite{PhysRevLett.51.605,PhysRevB.50.17199}.
	The Laughlin states are the \emph{exact} zero-energy ground states of $H$ provided that the interaction is sufficiently short-ranged~\cite{PhysRevB.31.5280}.

	By introducing $c_n^\dagger$ ($c_n$) as the creation (annihilation) operator for the orbital $|n\rangle$, the pseudopotential Hamiltonian can be recast in the separable form~\cite{PhysRevLett.92.096401,PhysRevLett.95.266405,PhysRevLett.107.126803,PhysRevB.88.035101,PhysRevB.88.165303}
	\begin{align}
	    H ={}& \sum_{m\ge 0} \mathcal U^m \Bigg(\sum_{j=0}^{N_\phi-1}{V_{j}^m}^\dagger V_{j}^m + \sum_{j=1/2}^{N_\phi-3/2} {W_j^m}^\dagger W_j^m\Bigg),
	    \label{eq:pp_ham}
	\end{align}
	where
	\begin{equation}
	V_j^m = \sum_l \zeta_{l}^m c_{j-l}c_{j+l}, \quad W_j^m = \sum_l \eta_{l}^m c_{j-l}c_{j+l},
	\end{equation}
	with $\zeta_{l}^m = \zeta_{j,l}^m$ and $\eta_{l}^m = \eta_{j,l}^m$ assumed to be independent of $j$, and with the convention that $c_n = c_n^\dagger = 0$ for $n < 0$ or $n \ge N_\phi $.
	The index $m$ is restricted to odd values for fermionic particles.
	Each term in $H$ describes a pair-hopping process from the orbitals $(j-l, j+l)$ to $(j-k, j+k)$, while conserving the total momentum $2j$.
	Here, $j$ takes integer values in $V_j^m$ and half-integer values in $W_j^m$, while $k$ and $l$ are constrained by the requirement that $j\pm k$ and $j\pm l$ be integers. 
	The coefficients $\zeta_l^m$ and $\eta_l^m$ are proportional to $\mathsf{H}_m(\sqrt 2 \kappa l) e^{-\kappa^2 l^2}$ where $\mathsf H_m$ is the $m$th Hermite polynomial.
	For other two-dimensional underlying surfaces, such as the sphere or the torus, the Hamiltonian retains the same form, with $\zeta_{l}^m$ and $\eta_l^m$ encoding the geometric information of the manifold~\cite{PhysRevB.88.035101,PhysRevB.88.165303}.

	Keeping only the terms with $m < M$ with $M$ an odd integer in Eq.~(\ref{eq:pp_ham}), the Laughlin state can be shown to be the ground state $|\psi_g\rangle$ of $H$ with the filling fraction $\nu = 1/M$, which is annihilated by both $V_j^m$ and $W_j^m$ for all $j$ and $m < M$. 
	Furthermore, if $d$-dimensional coefficient vectors $\boldsymbol \zeta^m = (\zeta_{1}^m, \zeta_{2}^m, \dots, \zeta_{d}^m) $ and $\boldsymbol \eta^m = (\eta_{1/2}^m, \eta_{3/2}^m, \dots, \eta_{(2d-1)/2}^m)$ with $0\le m<M$ and $d = (M-1)/2 $ are linearly independent, respectively,
	then the pseudopotential Hamiltonian has a unique zero-energy $\nu=1/M$ ground state~\cite{PhysRevB.88.165303,PhysRevB.91.085103}.
	In the following, we restrict the dimensions of $\boldsymbol \zeta^m$ and $\boldsymbol \eta^m$ to $d$ and $d+1$, respectively.
	
	A Laughlin-like state can be prepared as a steady state of the Lindbladian
	\begin{align}
	    \mathcal L(\rho) ={}& \sum_{j=0}^{N_\phi-1} \mathcal D[V_j^m](\rho) + \sum_{j=1/2}^{N_\phi-3/2} \mathcal D[W_j^m](\rho) \notag \\
	     + & \sum_{j=0}^{N_\mathrm{cell-1}}\mathcal D[Q_j](\rho)\label{eq:lindbladian}
	\end{align}
	where $N_\phi = M N_\mathrm{cell} - M + 1$,
	\begin{equation}
		Q_j = \sqrt{\gamma}c_{Mj}^\dagger P_j = \sqrt{\gamma} c_{Mj}^\dagger \prod_{\substack{k=Mj-(M-1)/2 \\ k\ne Mj}}^{Mj+(M-1)/2} (1 - n_k),
	\end{equation}
	and $\mathcal D[O](\rho) = O\rho O^\dagger - \frac12\{O^\dagger O, \rho\}$~\cite{sm}.
	The pair-loss operators $V_j^m$ and $W_j^m$ annihilate all states with filling faction less than $1/M$, which are eliminated from the steady-state manifold by the pump channels in the second line of the Lindbladian.
	$Q_j$ refills the orbital $Mj$ at rate $\gamma$ only when the $(M-1)$ neighbouring orbitals $\{Mj-(M-1)/2, Mj-(M-1)/2+1,\dots,Mj+(M-1)/2\}$ are unoccupied, as enforced by the projector $P_j$.

	\emph{Dissipative preparation.}
		\begin{figure}[t]
		\centering
		\includegraphics[width=\linewidth]{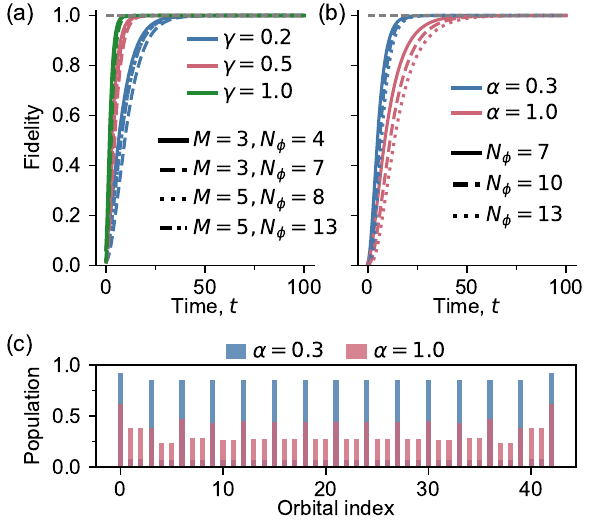}
		\caption{\quad Preparation of Laughlin-like states. 
		(a) Preparation of the TT state from the maximally mixed state for $\gamma=0.2$ and 0.5. 
		(b) Preparation of the Laughlin-like state Eq.~(\ref{eq:FQH_state}) from the maximally mixed state for $\gamma=0.5$ and $M=3$.
		(c) Particle-population distribution of the Laughlin-like state for $\alpha=0.3$ and 1 at $M=3$ and $N_\phi=43$.
		}\label{fig:preparation_fidelity}
	\end{figure}
	The Laughlin state can be expanded as $\sum_{\{n\}}C_{\{n\}}|\{n\}\rangle$ where $|\{n\}\rangle$ represents an occupation number eigenstate $|n_0,\dots, n_{N_\phi-1}\rangle$.
	Each component with nonzero $C_{\{n\}}$ can be inwardly squeezed from a root state called Tao-Thouless (TT) state $|\psi_\mathrm{TT}\rangle = |10\dots 010\dots 01\dots\rangle$~\cite{PhysRevB.28.1142,PhysRevB.50.17199,PhysRevLett.100.246802}, which is the ground state in the thin-torus limit, i.e., $L_1\rightarrow 0$.
	Since the TT state can be adiabatically connected to the Laughlin state in the limit $L_1\rightarrow \infty$~\cite{PhysRevB.28.2264,PhysRevLett.56.2395,PhysRevB.50.17199,PhysRevLett.94.026802,PhysRevLett.95.266405,Bergholtz2006,PhysRevB.77.155308,Jansen2008}, it serves as a natural target state for benchmarking the state preparation protocol. 
	In the TT state, any two adjacent particles occupy orbitals whose indices differ by $M-1$, satisfying a generalized $M$-Pauli principle~\cite{PhysRevLett.100.246802}.
	We therefore partition $N_\phi$ orbitals into $N_\mathrm{cell}$ cells, with $M$ orbitals in each bulk cell and $(M+1)/2$ orbitals in each boundary cell.
	The pump operators acting on the bulk (boundary) cells replenish one particle at the center (boundary) orbital of each cell only if the cell is empty.
	For states in the sector $N_e < N_\mathrm{cell}$, the pumping process continues until at least one orbital is occupied within each cell.
	On the contrary, for states in the sector $N_e > N_\mathrm{cell}$, no state can be annihilated by all loss operators simultaneously.
	Therefore, the TT state in the sector $N_e = N_\mathrm{cell}$ is the only one satisfying all constraints and constitutes the unique steady state.
	The coefficient vectors are chosen as $d$ or $(d+1)$-dimensional standard basis.
	Hereafter, we set the reference loss rate $\eta_{1/2}^1 = 1$ and normalize all other loss/pump rates accordingly.

	Starting from the maximally mixed state $\rho(0)=I/2^{N_\phi}$, we plot in Fig.~\ref{fig:preparation_fidelity}(a) the evolution of the fidelity $F = \langle\psi_\mathrm{TT}|\rho(t)|\psi_\mathrm{TT}\rangle$ with $M=3$ at $N_\phi = 7$ and $N_\phi=10$, and $M=5$ at $N_\phi=8$ and $N_\phi=13$, where $\rho(t)$ denotes the density operator of the system at time $t$.
	We find that the convergence time to the TT state is governed primarily by $\gamma$, rather than $M$ or the system size $N_\phi$.
	This is further corroborated by the Lindbladian gap $\Delta = -\mathrm{Re}(\lambda_1)$ since $\Delta$ sets the characteristic relaxation time $1/\Delta$.
	$\lambda_1$ is the eigenvalue of Eq.~(\ref{eq:lindbladian}) with the largest non-zero real part.
	The gap is precisely equal to $\gamma/2$ regardless of $N_\phi$ and $M$.
	Noting that each jump operator in Eq.~(\ref{eq:lindbladian}) always maps a Fock state $|s\rangle$ into another one $|s'\rangle$, successively acting random jump operators on an arbitrary Fock state leads to the TT state without revisiting any configuration.
	These transitions can be described by a directional acyclic graph which supplies a topological order between vertices by using Kahn's algorithm~\cite{Kahn1962}.
	Then the Lindbladian $\mathcal L$ can be rewritten as an upper-triangle matrix such that the gap is fully determined by the diagonal elements~\cite{sm}.

	Next, we consider the Laughlin states on the cylinder with larger $L_1$, necessitating longer range of hopping terms in $V_j^m$ or $W_j^m$.
	For simplicity, we consider $M=3$ and set $\boldsymbol\eta^1 = (1, \alpha)^\mathsf{T}$ with $\alpha \in (0,1]$.
	The pair-loss operators in $W_j^m$ are changed to $c_{j-1/2}c_{j+1/2} + \alpha c_{j-3/2}c_{j+3/2}$.
	The unnormalized steady state is given by~\cite{PhysRevLett.109.016401,PRXQuantum.1.020309} 
	\begin{equation}
		|\psi_\mathrm{ss}\rangle = \prod_j(1 - \alpha c_{j+1}^\dagger c_{j+2}^\dagger c_{j+3} c_{j})|\psi_\mathrm{TT}\rangle.\label{eq:FQH_state}
	\end{equation}
	As shown in Fig.~\ref{fig:preparation_fidelity}(c), increasing $\alpha$ leads to a more uniform particle distribution across the orbitals.
	The resulting state approaches the liquid phase for larger $\alpha$, as further indicated by the total correlation function~\cite{sm}.
	In addition, we verify that the steady state maintains a high overlap with the exact Laughlin state up to $L_1 \simeq 6$~\cite{sm}.
	Our protocol readily prepares $|\psi_\mathrm{ss}\rangle$ without modifying the pump channels, as shown in Fig.~\ref{fig:preparation_fidelity}(b).
	The squeezed component of the steady state is generated by the mixing term $(\alpha c_{j+1/2}^\dagger c_{j-1/2}^\dagger c_{j-3/2}c_{j+3/2} + \mathrm{h.c.})$ in the anti-commutator $\{{W_j^m}^\dagger W_j^m, \rho\}$ of the Lindbladian. 
	At $\alpha=0.3$, the preparation times are comparable for $N_\phi = 7$, 10, and 13, whereas at $\alpha=1$ they become distinguishable. 
	
	\begin{figure}[t]
		\centering
		\includegraphics[width=\linewidth]{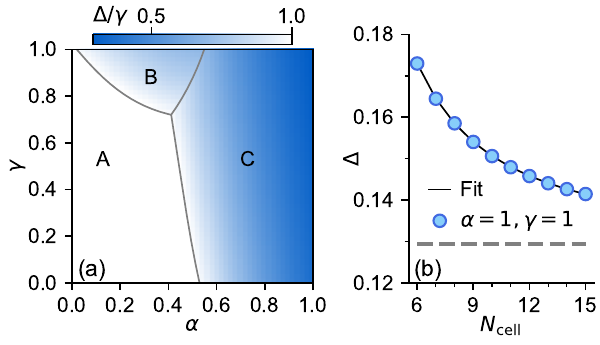}
		\caption{\quad Analysis of the Lindbladian gap $\Delta$. 
		(a) Two-dimensional map of the gap as a function of $\alpha$ and $\gamma$ over $0\le \alpha \le 1$ and $0.02 < \gamma\le 1$, for $M=3$ and $N_\mathrm{cell}=12$. 
		(b) Gap versus $N_\mathrm{cell}$ for $\alpha=1$ and $\gamma=1$ (blue circles).
		The solid black line is a fit to $\Delta_\infty + a / N_\mathrm{cell}^b$, with $\Delta_\infty=0.129$, $a=0.546$, and $b=1.411$.
		The dashed line indicates the asymptotic value $\Delta_\infty$.
		}
		\label{fig:gap}
	\end{figure}

	To estimate the preparation time of the Laughlin-like state with $0 < \alpha \le 1$ and $0 < \gamma \le 1$, we first determine the Lindbladian gap by exact diagonalization in the full doubled Hilbert space for $N_\mathrm{cell}=4$ with ten orbitals.
	We identify three candidate decay branches that control the gap in different parameter ranges.
	The branch-resolved analysis reproduces the full-space results throughout the entire parameter regime.
	We then extend the branch-resolved analysis to larger systems, up to $N_\mathrm{cell}=12$, and find the same gap structure with three regions $A$, $B$, and $C$, as shown in Fig.~\ref{fig:gap}(a)~\cite{sm}.

	In regions $A$ and $B$, the Lindbladian gap $\Delta$ is half of the gap of the effective Hamiltonian 
	\begin{align}
		H_\mathrm{eff} &= \sum_{j=0}^{N_\phi-2} n_jn_{j+1} + \sum_{j=0}^{N_\phi-3}n_j n_{j+2} + \alpha^2 \sum_{j=0}^{N_\phi-4} n_j n_{j+3} \notag \\
		&+\alpha\sum_{j=0}^{N_\phi-4}\left(c_{j+3}^\dagger c_j^\dagger c_{j+1}c_{j+2} + \mathrm{h.c.}\right) + \gamma\sum_{j=1}^{N_\mathrm{cell}}n_j^h.
	\end{align}
	with the corresponding eigenoperator $|\psi_\mathrm{ss}^{\mathrm{exc},(1)}\rangle\langle\psi_\mathrm{ss}|$ where $|\psi_\mathrm{ss}^{\mathrm{exc},(1)}\rangle$ is the first excited state of $H_\mathrm{eff}$.
	In region $A$, the gap is $\gamma/2$ with the corresponding eigenoperator $|\psi_\mathrm{ss}^{h, (1)}\rangle\langle\psi_\mathrm{ss}|$ where $|\psi_\mathrm{ss}^{h, (1)}\rangle$ denotes a local hole with one empty cell.
	The effective pump Hamiltonian $\gamma\sum_{j=1}^{N_\mathrm{cell}}n_j^h$ imposes a penalty $\gamma$ on each hole, which generalizes the result for $\alpha=0$.
	In region $B$, the effective Hamilton gap is set by the eigenenergy of a local excitation produced by an orbital shift from the steady-state pattern.
	The boundary between $A$ and $B$ is given by~\cite{sm} 
	\begin{equation}
		\gamma = \frac{2\alpha^4 + 5\alpha^2 + 4 - \alpha\sqrt{4\alpha^4 + 17\alpha^2 + 16}}{2(\alpha^2 + 2)}.
	\end{equation}

	In region $C$, the gap is governed by the Lindbladian restricted to an invariant subspace of the doubled Hilbert space, spanned by states reachable from the vacuum state through successive quantum jumps.
	Unlike in regions $A$ and $B$, the corresponding eigenoperator is not an outer product of two pure states, but rather a combination of diagonal and off-diagonal components.
	Figure~\ref{fig:gap}(b) shows the gap as a function of $N_\mathrm{cell}$ for $\alpha=1$ and $\gamma=1$, together with a fit to $\Delta(N_\mathrm{cell}) = \Delta_\infty + a / N_\mathrm{cell}^b$, which closely matches the data.
	We have also calculated the gap at other parameter points in region $C$, all of which are well fitted by the same functional form~\cite{sm}.
	The coefficient $a$ is nearly identical for points sharing the same $\gamma$, suggesting that it depends primarily on $\gamma$.
	The nonzero asymptotic value $\Delta_\infty$ implies efficient preparation of the Laughlin-like state from initial states within the invariant subspace, including the vacuum state.

	\emph{Adiabatic pumping.}
	\begin{figure}[t]
		\centering
		\includegraphics[width=\linewidth]{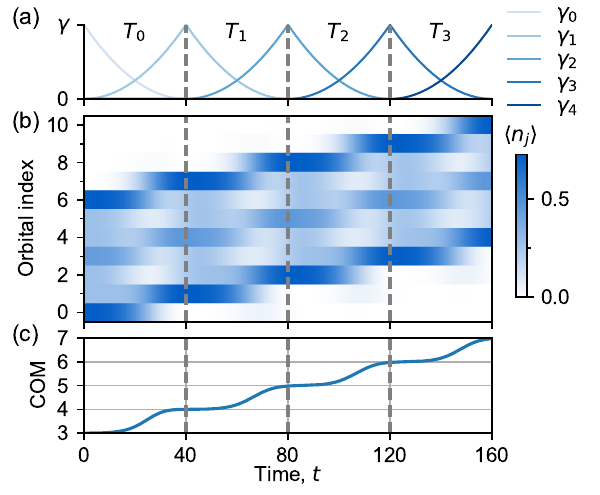}
		\caption{\quad Adiabatic pumping of a seven-orbital state.
		(a) Strengths of five pump tones as functions of time.
		(b) Adiabatic pumping dynamics of the particle population for a seven-orbital Laughlin-like state at $\alpha=0.8$.
		(c) Evolution of the center-of-mass (COM) of the state.
		}
		\label{fig:pump}
	\end{figure}
	Inspired by Laughlin's adiabatic charge pump~\cite{PhysRevB.23.5632}, we adiabatically pump the stabilized Laughlin-like state along a path that preserves the steady state.
	The target state, occupying seven orbitals, is pumped to the right by translating the pump operators in Eq.~(\ref{eq:lindbladian}) by one orbital per period $T$.
	Within each period, the pump strength $\gamma_i(t)$ at the trailing orbitals is ramped down to zero while a new, one-orbital-shifted pump strength $\gamma_{i+1}(t)$ is simultaneously ramped up from zero.
	Here $\gamma_0(t)$ denotes the strength of the pump stabilizing the initial state.
	Both strengths follow a quadratic ramp, $\gamma_i(t) = \gamma( 1 - t/T)^2$ and $\gamma_{i+1}(t) = \gamma (t/T)^2$, as shown in Fig.~\ref{fig:pump}(a).
	To ensure adiabaticity, we choose $T \gg 1/\Delta_\mathrm{min}$, where $\Delta_\mathrm{min}$ is the minimum instantaneous gap along the trajectory.
	The quadratic form helps reduce the boundary-induced effects~\cite{sm}.

	The initial state at $\alpha=0.8$, spanning orbitals zero to six, is adiabatically pumped to orbitals four to ten after four periods, as shown in Fig.~\ref{fig:pump}(b).
	The two-particle loss is imposed on the chain of eleven orbitals throughout the entire process.
	Since the pump only acts on a segment of orbitals, the steady-state manifold of the Linbladian on eleven orbitals is degenerate, and the final state cannot be prepared by directly quenching the pump to its final configuration.
	Figure~\ref{fig:pump}(c) depicts the evolution of center-of-mass (COM) $X = \sum_{j=0}^{N_\phi-1} \langle n_j\rangle j$ of the state, which starts at orbital three and ends at orbital seven.

	\emph{Experimental realization.}
	We propose an implementation of two-particle loss for spin-polarized fermions in a frozen optical-tweezer array by coupling occupied atom pairs to lossy Rydberg-macrodimer modes.  
	A global optical association tone with detuning $\varepsilon_\alpha$ couples a pair of ground-state atoms at separation $R$ to a macrodimer vibrational state whenever the two-photon resonance condition
	$\delta_\alpha(R)=2\varepsilon_\alpha+U_{\nu_\alpha}(R)/\hbar\simeq0$
	is satisfied, where $U_{\nu_\alpha}(R)$ is the selected Rydberg-pair molecular potential.
	Thus, in a one-dimensional array with spacing $s$, different tones can address different distance classes, e.g. $R=s$ and $R=2s$, yielding after elimination of the lossy molecular modes the dissipator
	$\dot\rho=\sum_j\Gamma_1{\mathcal D}[c_jc_{j+1}](\rho)+\sum_j\Gamma_2{\mathcal D}[c_jc_{j+2}](\rho)$.
	The required ingredients are motivated by direct quantum-gas-microscope observations of Rydberg macrodimers through correlated atom loss, distance-selective macrodimer dressing, and programmable fermionic tweezer arrays~\cite{Hollerith2019,Hollerith2022,Spar2022,Yan2022,Hollerith2024}.  
	Long-range jump operators can be engineered by coupling two distinct macrodimer modes, corresponding to different distance classes, to a common short-lived state.
	The which-path information is thereby erased by adiabatic elimination, and the obtained jump operator is collective, for example, $L_j\propto c_{j+1}c_{j+2}+c_{j}c_{j+3}$~\cite{sm}.

	The same tweezer platform also allows a three-site projected pump.  
	A source tweezer, continuously reloaded from an atomic reservoir and kept close to unit occupation, is placed near the middle target site and coupled to it by a Raman-assisted injection tone.  
	The three target tweezers are simultaneously off-resonantly dressed to a Rydberg state, generating an occupation-dependent addition energy for the middle site, $E_i^{\mathrm{add}} = E_i^0+ V_{\mathrm{dress}}(s)(n_{i-1}+n_{i+1})$.
	The injection tone is tuned to the unblocked transition at energy $E_i^0$, so that the middle site is filled only when both neighbouring sites are empty.  
	If either neighbour is occupied, the Rydberg-dressed interaction shifts the injection transition out of resonance and suppresses the pump.

	\emph{Conclusion and discussion.}
	We construct a Lindbladian with local loss and pump channels that stabilize a Laughlin-like state as the unique steady state under open boundary conditions.
	By identifying the two edge orbitals, the same construction yields an $M$-dimensional steady-state space on a torus, which is spanned by the original state and its $(M-1)$ translated copies~\cite{PhysRevLett.55.2095,PhysRevB.41.9377,Santos2020}.
	The protocol naturally extends to other FQH states.
	In particular, bosonic $\nu=1/2$ Laughlin-like states can be prepared within the framework by introducing additional on-site two-particle loss, a technique already demonstrated experimentally on superconducting circuits~\cite{Leghtas2015,PhysRevX.8.021005,Lescanne2020,Rglade2024} and neutral atoms~\cite{Syassen2008,Tomita2017}, together with the overlapping conditional pumps described in the Supplemental Material~\cite{sm}.
	More generally, related constructions may be applicable to unprojected Jain states whose parent Hamiltonians admit a separable form~\cite{PhysRevLett.124.196803}.
	Our results establish tunable local dissipation as a versatile tool for preparing and exploring topologically ordered phases of matter in current quantum systems.

	\begin{acknowledgments}
	We thank Zhao Liu for fruitful discussions on FQH states.
	This research was financially supported by the Knut and Alice Wallenberg Foundation through the Wallenberg Center for Quantum Technology (WACQT).
	\end{acknowledgments}

	% \bibliography{ref}

\end{document}

\title{Supplemental Material for ``Dissipative preparation of Laughlin-like states''}
	\author{Tong Liu}
	\affiliation{Chalmers Next Labs, 41296 Gothenburg, Sweden}

	\maketitle

	\section{Total correlation function}
	\begin{figure}[h]
		\centering
		\includegraphics[width=0.7\linewidth]{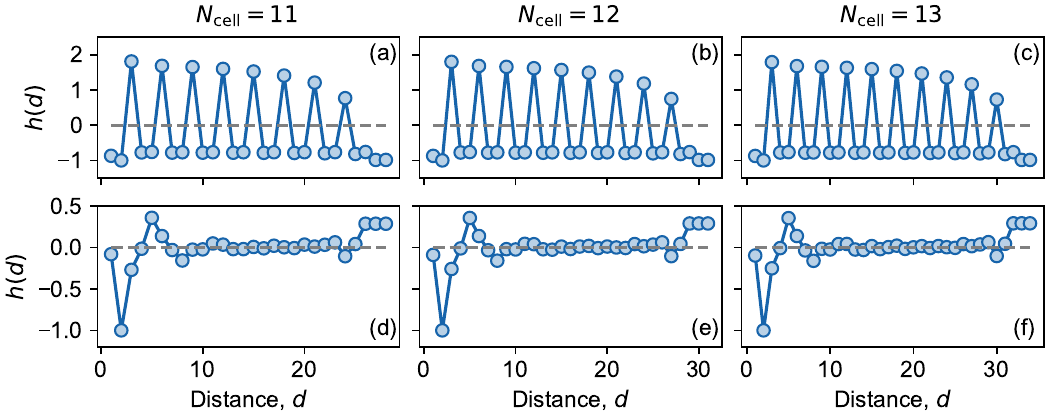}
		\caption{\quad Total correlation function of Laughlin-like states with $\nu=1/3$. 
		(a)-(c) $\alpha=0.2$. (d)-(f) $\alpha=1$.}
		\label{fig:hd_correlation}
	\end{figure}

	Figure~\ref{fig:hd_correlation} shows the total correlation function $h(d) = g(d) - 1$ as a function of the separation distance $d$ in the $\nu=1/3$ Laughlin-like state.
	Results for $\alpha=0.2$ and $\alpha=1$ are shown in the first and the second rows, respectively.
	We define the radial distribution function $g(d)$ as
	\begin{equation}
		g(d) = \frac{1}{N_d}\frac{\sum_j^{N_d} \langle n_j n_{j+d}\rangle}{\bar n^2},\quad \bar n = \frac{1}{N}\sum_j \langle n_j\rangle,
	\end{equation}
	where $N_d$ is the number of orbital pairs separated by $d$.
	To mitigate boundary effects, the first and last orbitals are excluded from the evaluation of both $g(d)$ and $\bar n$.

	For $\alpha=0.2$, $h(d)$ displays pronounced oscillations with period three, indicating a strong crystalline order.
	In contrast, for $\alpha=1$, $h(d)$ rapidly decays to zero in the bulk, consistent with liquid-like correlations.
	The weak upturn at large $d$ originates from residual boundary effects when the separation becomes comparable to the total length.

	\section{Overlap between the Laughlin-like states and Laughlin states}
	\begin{figure}[h]
		\centering
		\includegraphics[width=0.4\linewidth]{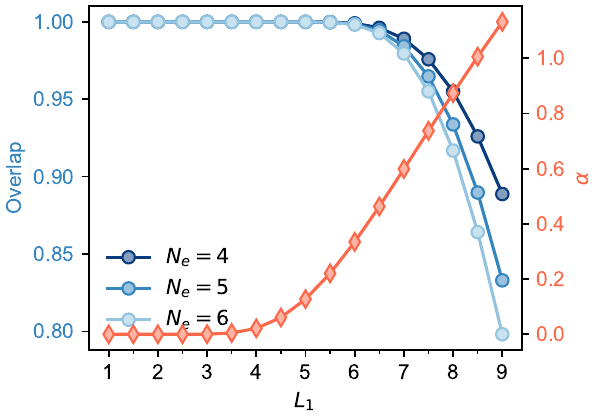}
		\caption{\quad Overlap between $\nu=1/3$ Laughlin-like states and Laughlin states as a function of $L_1$.}
		\label{fig:overlap}
	\end{figure}

	Figure~\ref{fig:overlap} shows the overlap $|\langle\psi_L|\psi_\mathrm{ss}\rangle|^2$ between the $\nu=1/3$ Laughlin-like state $|\psi_\mathrm{ss}\rangle$ in Eq.~(5) of the main text and the exact Laughlin state $|\psi_L\rangle$ on the cylinder with $N_e=4$, 5, and 6. 
	Using $\alpha$ as a variational parameter, we find the maximum overlap at $\alpha=3e^{-2\kappa^2}$ with $\kappa=2\pi/L_1$. 
	The overlap remains close to unity up to $L_1\simeq 6$. 
	Moreover, throughout $0\le\alpha\le 1$, the Laughlin-like state maintains a high overlap with the exact Laughlin state.

	\section{Lindbladian gap}

	As in the main text, we focus on the Lindbladian gap with $M=3$ in the parameter regime $0\le \alpha \le 1$ and $0 < \gamma \le 1$.

	\subsection{Tao-Thouless limit: $\alpha = 0$}
	In the Tao-Thouless (TT) limit, each jump operator $V_j^m$ or $W_j^m$ only involves a two-particle loss operator $c_i c_j$.
	Along with the pump operator, for any Fock state $|s\rangle$, each jump operator $L_k$ sends it to another Fock state $|t_k(s)\rangle$, i.e., $L_k|s\rangle = \phi_k(s)|t_k(s)\rangle$ if $\langle s|L_k^\dagger L_k|\rangle s\rangle = 1$ and $L_k|s\rangle = 0$ otherwise, where $\phi_k(s) = \pm 1$.
	Note that the recycling terms always change the particle number in double space by the same amount of particle $|s\rangle\langle s'| \rightarrow |s-2\rangle\langle s'-2|$ or $|s\rangle\langle s'| \rightarrow |s + 1\rangle\langle s'+1 |$ and the anti-commutator part preserves the particle number.
	The entire doubled Hilbert space can be split into two diagonal and off-diagonal subspaces.
	The eigenvalues of $\mathcal L$ is thus the union of the eigenvalues of two blocks.

	In the diagonal subspace $\mathcal X$ with $s = s'$, the Lindbladian acts as a classical Markov process generator.
	Let $\mathcal G$ be the directional jump graph on Fock states with an edge between $(s, t_k(s))$ for every $(k, s)$ with $ \langle s|L_k^\dagger L_k|s\rangle = 1$.
	Then $\mathcal G$ is acyclic and has no self-loops since pump operators can only replenish particles separated by $(M-1)$ orbitals while the loss operators deplete particles occupying at nearest-neighboring (NN) or next-nearest-neighboring (NNN) orbitals.
	By applying Kahn's algorithm for a directional acyclic graph, the Fock states $\{s\}$ can be relabeled as $s_1, s_2, \dots, s_{2^{N_\phi}}$ such that every transition $s_i \rightarrow s_j$ driven by a jump operator satisfies $i < j$.
	Therefore, $\LX$ can be rewritten as a lower-triangle matrix in this ordering.
	The eigenvalues are given by the diagonal elements which are contributed by the anti-commutator part $-\frac{1}{2}\left\{H_\mathrm{eff}, \cdot\right\}$ of the Lindbladian with
	\begin{equation}
		H_\mathrm{eff} = \sum_{j=0}^{N_\phi-2} n_j n_{j+1} + \sum_{j=0}^{N_\phi-3}n_j n_{j+2} + \gamma\sum_{j=1}^{N_\mathrm{cell}}n^h_j,\label{eq:diag_ham}
	\end{equation}
	where $n^h_j$ denotes the local hole number (0 or 1) at the cell $j$.
	$H_\mathrm{eff} \ge 0$ as each term in Eq.~(\ref{eq:diag_ham}) is a projector.
	The null space of $H_\mathrm{eff}$ is one-dimensional and spanned by the TT state.
	When $\gamma < 1$, the smallest non-zero eigenvalue of $H_\mathrm{eff}$ is $\gamma$, corresponding to one local hole in the TT state.
	Taking $N_\mathrm{cell}=3$ as an example, the null space only contains the TT state $|0, 3, 6\rangle$ but the one-local-hole subspace is ten-fold degenerate, spanned by the states $|0, 3\rangle$, $|1, 4\rangle$, $|2, 5\rangle$, $|2, 6\rangle$, $|3,6\rangle$, $|0, 4\rangle$, $|0, 5\rangle$, $|0, 6\rangle$, $|1,6\rangle$ and $|1, 5\rangle$.
	The largest non-zero eigenvalue of $\LX$ is thus $-\gamma$.

	In the off-diagonal subspace $\mathcal Y$, the Lindlbian acts on $|s\rangle\langle s'|$ $(s\ne s')$ as
	\begin{equation}
		\mathcal L(|s\rangle\langle s'|) = -\frac12\left(E(s) + E(s')\right)|s\rangle\langle s'| + \sum_k \phi_k(s)\phi_k(s')c_k(s)c_k(s')|t_k(s)\rangle\langle t_k(s')|,
		\label{eq:lin_off_diag}
	\end{equation}
	where $E(s) = \langle s|H_\mathrm{eff}|s\rangle$ and $c_k(s) = \langle s| L_k^\dagger L_k|s\rangle$.
	Define the doubled jump graph $\mathcal G^{(2)}$ on pairs $(s, s')$ with an edge $(s, s') \rightarrow (t_k(s), t_k(s'))$ for every $k$ with $c_k(s) c_k(s') = 1$ and $s \ne s'$.
	Similarly, the graph $\mathcal G^{(2)}$ is also acyclic because a cycle in $\mathcal G^{(2)}$ projects to a closed walk on the first coordinate in $\mathcal G$ which contradicts the acyclicity of $\mathcal G$.
	Therefore, the second term in Eq.~(\ref{eq:lin_off_diag}) is strictly lower-triangle after employing Kahn's algorithm to relabel the pairs $(s, s')$.
	The eigenvalues of $\LY$ is read off the diagonal:
	\begin{equation}
		-\frac12 \left(E(s) + E(s')\right),\quad s\ne s'.
	\end{equation}
	The largest non-zero eigenvalue is $-\gamma/2$ obtained when $\{E(s), E(s')\} = \{0, 1\}$.

	Combining the results of two subspaces, the gap $\Delta$ of Lindbladian is thus $\gamma/2$.

	\subsection{Beyond the Tao-Thouless limit: $0 < \alpha \le 1$}
	Inspired by the former case, we examine the excitation gap $\Delta^\mathrm{exc}$ of the effective Hamiltonian $H_\mathrm{eff} = \sum_k L_k^\dagger L_k$
	\begin{align}
	H_\mathrm{eff} &=  H_\mathrm{loss} + H^h, \notag \\
	 H_\mathrm{loss} &= \sum_{j=0}^{N_\phi- 2}n_j n_{j+1} + \sum_{j=0}^{N_\phi - 3}n_j n_{j+2} +  \alpha^2\sum_{j=0}^{N_\phi - 4}n_j n_{j+3} +  \alpha\sum_{j=0}^{N_\phi-4} \left( c_{j+3}^\dagger c_j^\dagger c_{j+1} c_{j+2} + \mathrm{h.c.}\right),\notag \\
	 H^h &= \gamma\sum_{j=1}^{N_\mathrm{cell}} n_j^h .
	\end{align}
	The steady state $|\psi_\mathrm{ss}\rangle$ of $H_\mathrm{eff}$ is given by Eq.~(5) in the main text.
	By denoting the first excited state of $H_\mathrm{eff}$ as $|\psi_\mathrm{ss}^{\mathrm{exc},(1)}\rangle$, $|\psi_\mathrm{ss}\rangle\langle\psi_\mathrm{ss}^{\mathrm{exc,(1)}}|$ is the eigenoperator of $\mathcal L$ since
	\begin{align}
		\mathcal L(|\psi_\mathrm{ss}\rangle\langle\psi_\mathrm{ss}^{\mathrm{exc}, (1)}|) ={}& \sum_{k} L_k |\psi_\mathrm{ss}\rangle\langle\psi_\mathrm{ss}^{\mathrm{exc}, (1)}|L_k^\dagger - \frac{1}{2}H_\mathrm{eff}|\psi_\mathrm{ss}\rangle\langle\psi_\mathrm{ss}^{\mathrm{exc}, (1)}| - 
		\frac12|\psi_\mathrm{ss}\rangle\langle\psi_\mathrm{ss}^{\mathrm{exc}, (1)}|H_\mathrm{eff} \notag \\
		={}&-\frac{\Delta^\mathrm{exc}}{2}|\psi_\mathrm{ss}\rangle\langle\psi_\mathrm{ss}^{\mathrm{exc}, (1)}|.
	\end{align}
	The cycling terms vanish because $L_k|\psi_\mathrm{ss}\rangle = 0$.
	Similarly, $|\psi_\mathrm{ss}^{\mathrm{exc},(1)}\rangle\langle\psi_\mathrm{ss}|$ is another eigenoperator of $\mathcal L$ with the same eigenvalue $-\Delta^{\mathrm{exc}} / 2$.
	As $\Delta^\mathrm{exc}$ controls the Lindbladian gap, we first examine the gap of the effective Hamiltonian.

	\subsection{The gap of the effective Hamiltonian}

	\begin{figure}[h]
		\centering
		\includegraphics[width=0.8\linewidth]{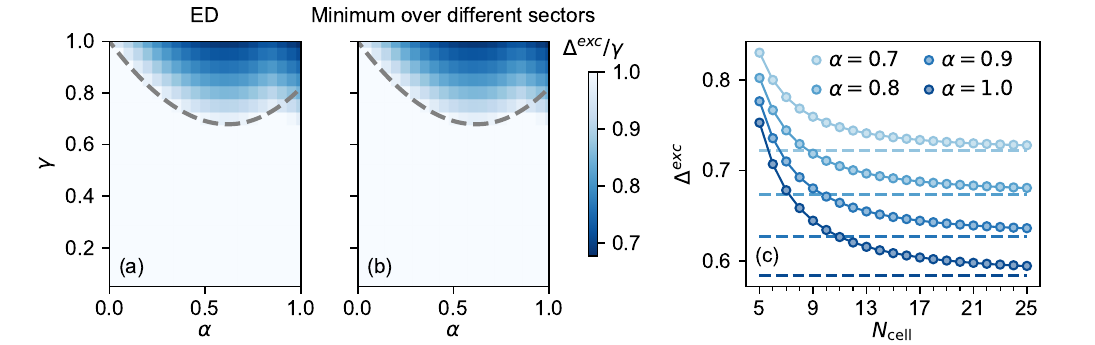}
		\caption{\quad Gap $\Delta^\mathrm{exc}$ of the effective Hamiltonian.(a) Gap as a function of $(\alpha, \gamma)$ for $N_s = 19$ by exact diagonalization in the entire Hilbert space.
		(b) Gap over one-local-hole sector, $\Delta K = 0$ sector, and $\Delta K = 1$ sector.
		The dash line is given by Eq.~(\ref{eq:boundary}).
		(c) Gap as a function of $N_\mathrm{cell}$ in the sector $\Delta K = 0$ with various $\alpha$.
		All can be well fitted by the function $\Delta_\infty^\mathrm{exc} + a/N_\mathrm{cell}^b$.
		Dashed lines indicate the asymptotic values.
		}
		\label{fig:gap_delta_exc}
	\end{figure}

	As shown in the Fig.~3(a) of the main text, the Lindbladian gap $\Delta$ is set by the lowest of three branches.
	We first consider the one-local-hole sector, in which the steady state contains one local hole.
	Since the local-hole states are also annihilated by the two-particle loss operators, only $H^h$ contributes to the excitation energy.
	Accordingly, the one-local-hole excitation energy is $\gamma$ and the gap $\Delta$ is given by $\gamma/2$.

	Since the effective Hamiltonian converses the total momentum, we consider the sector with the momentum change $\Delta K = 1$ with respect to the TT state.
	Starting from the root state $|0, 3, 5, \dots \rangle$, the minimal realization appears in the three-dimensional subspace spanned by the states $|0, 3, 5,\dots\rangle$, $|1, 2, 5,\dots\rangle$, $|1, 3, 4,\dots\rangle$, and is described by the Hamiltonian $H^\mathrm{defect}$
	\begin{equation}
		\begin{pmatrix}
		1 + \alpha^2 & 		\alpha  & 0		\\
		\alpha 		 & 1 + \alpha^2 & \alpha		 \\
		0 		 & 		 \alpha & 2 + \alpha^2 + \gamma 
		\end{pmatrix}.
	\end{equation}
	Requiring the smallest eigenvalue $\lambda_1$ to satisfy $\lambda_1 > \gamma$, we obtain
	\begin{equation}
		\gamma < \frac{2\alpha^4 +5\alpha^2 + 4 - \alpha\sqrt{4\alpha^4 + 17\alpha^2 + 16}}{2(\alpha^2 + 2)}.\label{eq:boundary}
	\end{equation}
	The local defect block is decoupled from the remaining part with two permanently empty buffer orbitals $7$ and $8$.
	No allowed four-site pattern $1001$ or $0110$ can cross this buffer.
	The full Hilbert space thus factorizes as $\mathcal H^\mathrm{defect}\otimes \mathcal H^\mathrm{tail}$, and the effective Hamiltonian decouples into $H^\mathrm{defect} + H_\mathrm{eff}^\mathrm{tail}$.
	Since $H_\mathrm{eff}^\mathrm{tail}$ is positive, the smallest eigenvalue of $H_\mathrm{eff}$ coincides with the smallest eigenvalue $\lambda_1$ of $H^\mathrm{defect}$.

	Figure~\ref{fig:gap_delta_exc}(a) shows the gap as a function of $\alpha$ and $\gamma$, obtained by exact diagonalization for $N_\mathrm{cell}=7$ and $N_s=19$.
	Dividing the gap by $\gamma$ makes the crossover between the $\gamma-$dominated and $\lambda_1$-dominated regions clearly visible.
	The boundary is given by Eq.~(\ref{eq:boundary}).

	\subsection{The gap in the sector $\Delta K = 0$}

	The upper-right corner of Fig.~\ref{fig:gap_delta_exc}(a), however, deviates from $\lambda_1$, indicating the presence of another type of excitation in the sector $\Delta K = 0$.
	In this sector, there are neither holes nor particle pairs separated by a distance of 2, so the gap is fully determined by
	\begin{equation}
		\sum_j n_j n_{j+1} + \alpha^2\sum_j n_jn_{j+3} + \alpha\sum_j\left(c_{j+3}^\dagger c_j^\dagger c_{j+1} c_{j+2} + \mathrm{h.c.}\right)\label{eq:eff_ham_k0} .
	\end{equation}
	Figure~\ref{fig:gap_delta_exc}(c) shows the gap in the sector $\Delta K=0$ as a function of $N_\mathrm{cell}$ for several values of $\alpha$.
	Each curve is well fitted by the function $\Delta^\mathrm{exc} = \Delta_\infty^\mathrm{exc} + a / N_\mathrm{cell}^b$ with fitting variables $(\Delta_\infty^\mathrm{exc}, a, b)$.

	The gap can be explained by the equivalent Glauber dynamics.
	We define an effective variable $x_k \in \{0, 1\}$, $k = 1, \dots, N_\mathrm{cell}-1$, where $x_k = 1$ means the $k$-th local pattern has jumped: $1001\rightarrow 0110$.
	Two neighboring jumps cannot both occur, so $x_k x_{k+1}=0$.
	Thus the effective configurations are exactly
	\begin{equation}
		\Omega = \{x\in \{0, 1\}^{N_\mathrm{cell}-1}: x_k x_{k+1} = 0\}.
	\end{equation}
	These are the independent sets of the path graph $\mathcal G$.
	The condition $x_k x_{k+1} = 0$ is the hard-core constraint.
	We define $|x| = \sum_k x_k$ which counts the number of effective jumps performed.
	Let $A$ be the adjacency matrix of the one-jump between configurations in the graph.
	$A(x,y) = 1$ if and only if $x$ and $y$ differ by flipping one $x_k$, and $x$ and $y$ are allowed.
	Otherwise $A(x, y) = 0$.
	For any $x \in \Omega$, define
	\begin{equation}
		C_1(x) = \#\{k: x_k=1\}, \quad C_0(x) = \#\{k: x_k = 0, \ x_{k-1} = 0, \ x_{k+1} = 0\}.
	\end{equation}
	Here $C_1(x)$ counts allowed annihilation moves $1\to 0$, and $C_0(x)$ counts allowed creation moves $0 \to 1$.
	Then the Hamiltonian in Eq.~(\ref{eq:eff_ham_k0}) can be expressed as 
	\begin{equation}
		H_\alpha(x,y)=
			\begin{cases}
			W_\alpha(x) = C_1(x) + \alpha^2 C_0(x),&x=y,\\
			\alpha, &x \sim y,\\
			0,&\text{otherwise},
			\end{cases}
	\end{equation}
	since a local $0110$ pattern contributes $1$, while a local $1001$ pattern contributes $\alpha^2$.
	$x\sim y$ means that $x$ and $y$ are connected by one allowed effective flip.
	For simplicity, we define the sign-transformed matrix 
	\begin{equation}
	M_\alpha = U H_\alpha U^{-1} = W_\alpha - \alpha A,
	\end{equation}
	where $U|x\rangle = (-1)^{|x|}|x\rangle$.
	By defining the projector $P_k$ at the site $k$ $P_k = |0_k\rangle\langle 0_k|$, $M_\alpha$ can be rewritten as
	\begin{equation}
		M_\alpha = \sum_k P_{k-1} \begin{pmatrix}
			\alpha^2 & -\alpha \\
			-\alpha &  1 \\
		\end{pmatrix}
		 P_{k+1}.
	\end{equation}
	The $2\times 2$ matrix is positive semidefinite as
	\begin{equation}
		\begin{pmatrix}
			\alpha^2 & -\alpha \\
			-\alpha & 1 \\
		\end{pmatrix}
		 = \begin{pmatrix}
		 	\alpha \\ -1
		 \end{pmatrix} \begin{pmatrix}
		 	\alpha & -1
		 \end{pmatrix}.
	\end{equation}
	The zero-energy ground state is 
	\begin{equation}
	 	|\Psi_\alpha\rangle = \sum_{x\in \Omega} \alpha^{|x|}|x\rangle.
	\end{equation}

	The hard-core Gibbs measure with fugacity $\lambda > 0$ on the path $\mathcal G$ is the probability measure on independent sets
	\begin{equation}
		\mu_\lambda(x) = \frac{\lambda^{|x|}}{Z(\lambda)},\quad Z(\lambda) = \sum_{x\in \Omega}\lambda^{|x|}.
	\end{equation}
	In our case we set $\lambda = \alpha^2$.
	The ground-state amplitudes are proportional to $\sqrt{\mu_{\alpha^2}(x)}$.
	We can show that the measure $\mu_{\alpha^2}$ can be generated by a heat-bath Glauber generator.
	If one of the neighbors of $k$ is occupied, then $x_k = 0$ forced by the hard-core constraint.
	If both neighbors are empty, then set $x_k = 1$ with probability $\alpha^2/(1 + \alpha^2)$ and set $x_k = 0$ with probability $1/(1+\alpha^2)$.
	The continuous-time generator $Q$ is
	\begin{equation}
		Q(x, y) = 
		\begin{cases}
			\dfrac{\alpha^2}{1 + \alpha^2}, & y = x^k, \ x_k = 0, \ x_{k-1} = x_{k+1} = 0, \\[1.2em]
			\dfrac{1}{1 + \alpha^2}, & y = x^k, \ x_k = 1, \\[1.2em]
			-\displaystyle\sum_{z\ne x} Q(x, z), & y = x, \\ 
			0, & \text{otherwise.}
		\end{cases}
	\end{equation}
	Here again $x_0 = x_{N_\mathrm{Ncell}} = 0$ at the boundaries.
	Thus $Q$ is exactly the heat-bath Glauber generator for the hard-core Gibbs measure $\mu_{\alpha^2}$.
	It satisfies detailed balance:
	\begin{equation}
		\mu_{\alpha^2}(x)Q(x, y) = \mu_{\alpha^2}(y)Q(y, x).
	\end{equation}
	For example, if $y = x^k$ is obtained from $x$ by a $0\to 1$ flip, then $\mu_{\alpha^2}(y)/\mu_{\alpha^2}(x) = \alpha^2$, while
	\begin{equation}
		\frac{Q(x, y)}{Q(y, x)} = \frac{\alpha^2/(1+\alpha^2)}{1/(1 + \alpha^2)} = \alpha^2.
	\end{equation}
	Therefore, the detailed balance condition holds.

	To connect the generator $Q$ with the matrix $M_\alpha$, we show that
	\begin{equation}
	Q = -\frac{1}{1 + \alpha^2} D_\mu^{-1/2}M_\alpha D_\mu^{1/2},\quad D_u(x, x) = \mu_{\alpha^2}(x).\label{eq:sym_Q}
	\end{equation}
	$Q(x,y)$ and $M_\alpha(x,y)$ are both nonzero only when $x\sim y$.
	Suppose $y = x^k$ with $x_k = 0$ and $y_k=1$.
	Substituting $M_\alpha(x, y) = -\alpha$ and $\sqrt{\mu_{\alpha^2}(y)/\mu_{\alpha^2}(x)} = \alpha$ into Eq.~(\ref{eq:sym_Q}) yields
	\begin{equation}
	Q(x, y) = \frac{\alpha^2}{ 1 + \alpha^2}, 
	\end{equation}
	which coincides with the definition of $Q$.
	Similarly, we can also verify that Eq.~(\ref{eq:sym_Q}) holds for $x_k=1$ and $y_k=0$.
	For $y=x$, we have
	\begin{equation}
		Q(x,x) = -\frac{C_1(x) + \alpha^2 C_0(x)}{1+\alpha^2}.
	\end{equation}
	which is exactly the sum of all outgoing rates from $x$, $\sum_{y\ne x} Q(x, y)$.
	We now consider the spectrum of the generator $Q$ since $Q$ and $M_\alpha$ share the same spectrum according to Eq.~(\ref{eq:sym_Q}). 

	Let $K_k$ be the local update kernel at site $k$.
	A discrete-time random-scan Glauber transition matrix is given by
	\begin{equation}
		P_\mathrm{disc} = \frac{1}{N_\mathrm{cell} - 1}\sum_k K_k.
	\end{equation}
	It means that one site is uniformly picked at random, and is updated by the local kernel $K_k$.
	The continuous-time generator $Q$ is defined as
	\begin{equation}
		Q = \lim_{dt\rightarrow 0}\frac{P_{dt} - 1}{dt},\quad P_t(x,y) = \mathrm{Pr}(X_t=y|X_0=x).
	\end{equation}
	Because the local sites are independent, the transition probability during a short duration $dt$ is
	\begin{equation}
		P_{dt} = (1 - (N_\mathrm{cell}-1)dt)I + \sum_k K_k dt + O(dt^2) = I + \sum_{k}(K_k - I)dt + O(dt^2).
	\end{equation}
	Therefore
	\begin{equation}
		Q = \sum_{k=1}^{N_\mathrm{cell}-1}(K_k - I) = (N_\mathrm{cell}-1)(P_\mathrm{disc} - I).
	\end{equation}
	The continuous-time spectral gap is $(N_\mathrm{cell}-1)$ times the discrete-time spectral gap.

	The recent result of Glauber dynamics states that~\cite{Chen2024}, for antiferromagnetic two-spin systems in the tree-uniqueness regime, the discrete-time Glauber dynamics has an optimal $\Omega(n^{-1})$ lower bound on the spectral gap; for the hard-core model with fugacity $\lambda$ and maximum degree $D$, one sufficient condition is
	\begin{equation}
	\lambda\le(1-\delta)\lambda_c(D),
	\qquad
	\lambda_c(D)=\frac{(D-1)^{D-1}}{(D-2)^D},
	\label{eq:uniqueness-condition}
	\end{equation}
	for some fixed $\delta\in(0,1)$.  

	Our effective graph is a path, so it has maximum degree $2$.  
	To use the standard theorem stated for maximum degree at most $D\ge3$, we may embed the path class into the class of graphs with maximum degree at most $3$.  
	Then the sufficient condition becomes $\lambda < 4$, which is equivalent to $\alpha < 2$.
	For such fixed $\alpha$, choose $\delta=1-\frac{\alpha^2}{4}>0$.
	The theorem gives a constant $c(\alpha)>0$, independent of the system size, such that
	\begin{equation}
	\mathrm{gap}_{\mathrm{disc}}\ge \frac{c(\alpha)}{N_\mathrm{cell}-1}
	\end{equation}
	We thus prove that the gap of the generator $Q$ is greater than $c(\alpha)$ for every fixed $0<\alpha<2$.

	From another perspective, the Hamiltonian in Eq.~(\ref{eq:eff_ham_k0}) is the parent Hamiltonian of an injective matrix product state (MPS).
	The MPS for the ground state has bond dimension 2,
	\begin{equation}
		M^{(r)} = \begin{pmatrix}
			1 & 0 \\
			1 & 0
		\end{pmatrix},\quad M^{(d)} = \begin{pmatrix}
			0 & -\alpha \\
			0 & 0
		\end{pmatrix},
	\end{equation}
	where each bond is either a rod state $|r\rangle = |1001\rangle$ or a dimer state $|d\rangle = |0110\rangle$.
	Since the products of these matrices span all $2\times 2$ matrices, the MPS is injective.
	A frustration-free parent Hamiltonian of an inject MPS has a unique ground state and a spectral gap bounded below by a positive constant independent of system size~\cite{Fannes1992}.
	Hence $H$ is gapped in the sector $\Delta K = 0$.

	Figure~\ref{fig:gap_delta_exc}(b) shows the minimum of the three gaps in the sectosr $\Delta K = 0$ and $\Delta K = 1$, together with $\gamma$, which agrees with the ED result, confirming that these three gaps entirely capture the spectral gap throughout the parameter range $0 \le \alpha \le 1$ and $0<\gamma\le 1$.

	\subsection{The gap of $\mathcal L$ in the invariant subspace}

	\begin{figure}[h]
		\centering
		\includegraphics[width=0.8\linewidth]{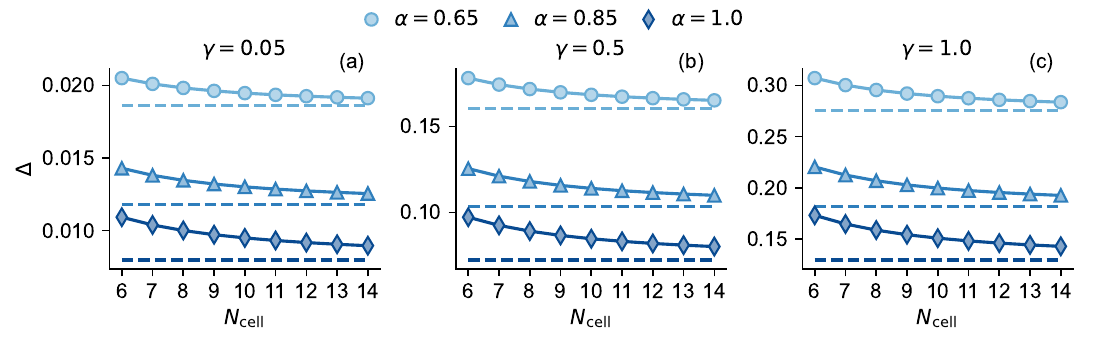}
		\caption{\quad Lindbladian gap in the invariant subspace.}
		\label{fig:reduced_gap}
	\end{figure}

	\begin{table}[h]
		\centering
		\begin{tabular}{l|ccc}
		\hline
						& $\alpha=0.65$ & $\alpha=0.85$ & $\alpha=1$ \\
		\hline
		$\gamma=0.05$	& (0.0186, 0.0294, 1.528) & (0.0118, 0.0305, 1.393) & (0.00798 , 0.0299, 1.295) \\
		$\gamma=0.5$    & (0.161, 0.290, 1.562) & (0.103, 0.295, 1.449) & (0.0723, 0.289, 1.368) \\
		$\gamma=1.0$	& (0.275, 0.540, 1.577) & (0.181, 0.548, 1.474) & (0.129, 0.546, 1.411) \\
		\hline
		\end{tabular}
		\caption{\quad Fitting coefficients used in Fig.~\ref{fig:reduced_gap}.}
		\label{tab:fit_para}
	\end{table}

	Starting from the vacuum state $|0\dots0\rangle\langle 0\dots 0|$, the system traverses a subspace through a sequence of jumps induced by $\mathcal L$.
	During this process, no two particles ever occupy orbitals separated by a distance of two since particles are only replenished in the orbitals $3j$ with $j = 0, 1, 2, \dots, N_\mathrm{cell}-1$.  
	Therefore, the gap within this subspace is independent of the jump operators $c_j c_{j+2}$, which can be removed from the Lindbladian.
	We generate all such states by repeatedly applying the Lindbladian and then diagonalize it in the basis of doubled Fock states spanning these states.

\begin{figure}
		\centering
		\includegraphics[width=0.6\linewidth]{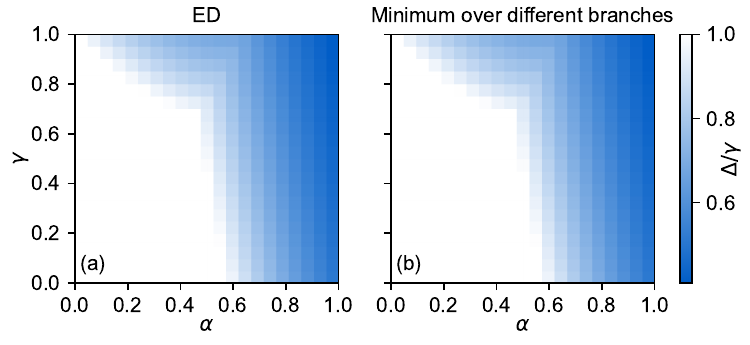}
		\caption{\quad Lindbladian gap for $N_\mathrm{cell}=4$.(a) Exact diagonalization in the full doubled Hilbert space.
		(b) Minimum over different branches.}
		\label{fig:gap_lindbladian}
	\end{figure}

	Figure~\ref{fig:reduced_gap} shows the Lindbladian gap in the invariant subspace as a function of $N_\mathrm{cell}$ for several values of $(\alpha, \gamma)$.
	Each data series is also well fitted by $\Delta_\infty + a / N_\mathrm{Ncell}^b$ with the fitting coefficients $(\Delta_\infty, a, b)$ reported in Table~\ref{tab:fit_para}.
	As shown in each row of Table~\ref{tab:fit_para}, the coefficient $a$ strongly depends on $\gamma$ and is essentially independent of $\alpha$, while the exponent $b$ lies in the range $[1, 2]$.
	Taking the mininum of the Lindbladian gap in the invariant subspace and $\Delta^\mathrm{exc}/2$ yields Fig.~(3)a in the main text.
	Figure~\ref{fig:gap_lindbladian} compares the gap obtained by exact diagonalization with the minimum over all branches.
	The two results agree to within numerical precision in the regions $0\le \alpha \le 1$ and $0 < \gamma \le 1$.
	We note that $\Delta^\mathrm{exc}$ in the sector $\Delta K = 0$ exceeds twice the Lindbladian gap in the restricted subspace in the right upper corner of the diagram ($0.6 < \alpha \le 1$ and $0.5 \le \gamma \le 1$).
	The Lindbladian gap is thus governed by the effective Hamiltonian gap $\Delta^\mathrm{exc}$ only in regions $A$ and $B$, as shown in Fig.~3(a) of the main text.

	\section{Adiabatic pump}

\begin{figure}[ht]
		\centering
		\includegraphics[width=0.5\linewidth]{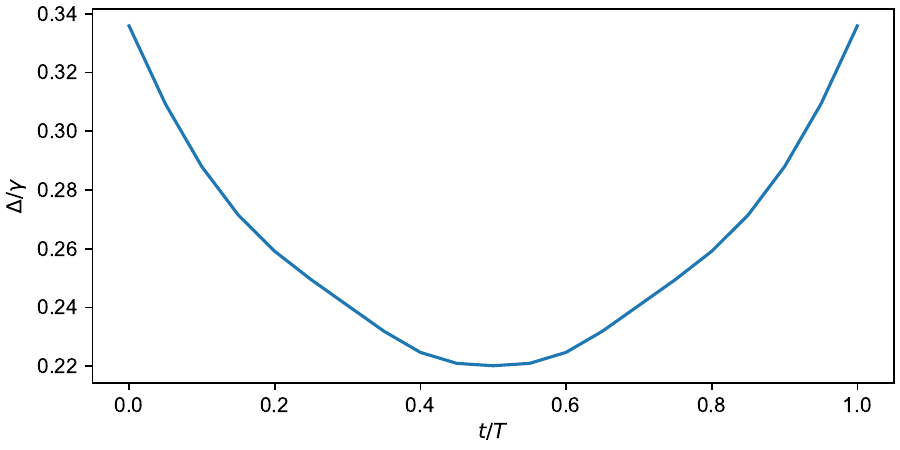}
		\caption{\quad The Lindbladian gap $\Delta$ within a pump period.}
		\label{fig:gap_pump}
	\end{figure}

	The Lindbladian gap $\Delta(t)$ over one period is shown as a function of $t/T$ in Fig.~\ref{fig:gap_pump}.
	It reaches the minimum $\Delta/\gamma\approx 0.22$ at $t/T=0.5$, coinciding with the minimum point of the scale factor $(1-t/T)^2 + (t/T)^2$. 
	For the maximal pump strength $\gamma=1$, we choose $T = 40$ to ensure $T \gg 1/\Delta$.

	The boundary contribution to the adiabatic error can be estimated from the rate of change of $\mathcal L_i(t)$ at the endpoint $t=T$ in the $i$-th period,
	\begin{equation}
		c_m = \mathrm{Tr}\left[\sigma_m^\dagger \partial_t \mathcal L_i(t)(\rho_\mathrm{ss})\right]\big\vert_{t=T}
	\end{equation}
	where $\rho_\mathrm{ss}$ is the steady state of $\mathcal L_i(t)$ at $t=T$ and $\sigma_m$ is the $m$-th left eigenoperator.
	For the quadratic path, the derivative is given by
	\begin{equation}
		\partial_t \mathcal L_i(t)\vert_{t=T} = \mathcal \mathcal \mathcal Q_{i+1},
	\end{equation}
	where $\mathcal Q_{i+1}$ is the pump dissipator at $t=T$ of the $i$-th period.
	Since $Q_{i+1}$ annihilates $\rho_\mathrm{ss}$, we obtain $c_m = 0$.

	\section{Preparation of $\nu=1/2$ Laughlin-like states}

	The target $\nu=1/2$ Laughlin-like states with $N_\phi = 2N_e - 1$ is expressed as
	\begin{equation}
		|\psi\rangle = \prod_j \left(1 - \frac{\alpha}{2}a_{j-1}a_{j+1}{a_j^{\dagger}}^2\right)|\psi_\mathrm{TT}\rangle,\label{eq:bosonic_FQH}
	\end{equation}
	where $a_j$ ($a_j^\dagger$) is the bosonic annihilation (creation) operator at the orbital $j$, and the corresponding TT state is $|1010\cdots 101\rangle$.
	The inward squeezing operator moves two adjacent bosons toward the central orbital, creating a doublon.

	For fermionic states, the two-particle loss operators $V_j^m$ start from $\zeta_j^1 c_{j-1}c_{j+1}$, whereas for bosonic states they start from $\zeta_j^0 a_j^2$, since two bosons can occupy the same orbital. 
	Moreover, in the bosonic case, squeezing occurs only between orbitals separated by an even number of orbitals.
	The squeezed states are thus induced by the associated loss operators $V_j^m = \zeta_j^0 a_j^2 + \alpha \zeta_j^1 a_{j-1}a_{j+1}$ rather than by $W_j^m$ as in the fermionic case.

	The pump operators take the same form as in the fermionic case, but overlap between neighboring operators.
	Taking $N_e=3$ and $N_\phi=5$ as an example, the pump operators include $a_0^\dagger P_1$, $P_1a_2^\dagger P_3$, $P_3 a_4^\dagger$ where $P_i$ is the projector onto the Fock state $|0_i\rangle$.

\begin{figure}[h]
		\centering
		\includegraphics[width=0.4\linewidth]{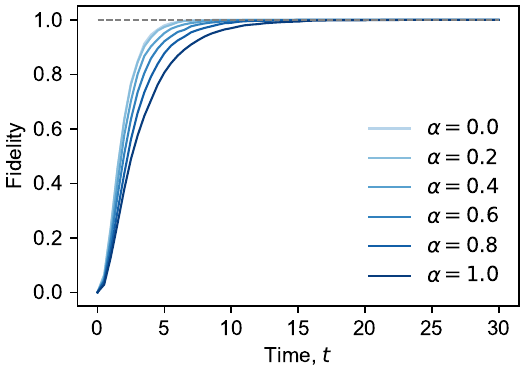}
		\caption{\quad Dissipative preparation of $\nu=1/2$ bosonic states from the vacuum state with $\gamma=1.0$ and $N_\phi=5$.}
		\label{fig:bosonic_prep}
	\end{figure}

	Figure~\ref{fig:bosonic_prep} shows the fidelity between the target state Eq.~(\ref{eq:bosonic_FQH}) and the evolved state as a function of time, starting from the vacuum state.
	States with larger $\alpha$ take longer to reach unit fidelity, similar to the fermionic case.

	\section{Experimental realization}

	\subsection{Physical setting and notation}

	We consider a one-dimensional array of frozen target tweezers with lattice spacing $s$ and spin-polarized fermions. 
	We denote by $c_j^\dagger$ the creation operator for a spin-polarized fermionic atom in the selected ground-state motional orbital 
	of tweezer $j$.
	The corresponding operators satisfy $\{c_i,c_j^\dagger\}=\delta_{ij}$.
	The target tweezers are assumed sufficiently deep and well separated that coherent tunnelling between target sites is negligible during the dissipative evolution.
	The central auxiliary object is a Rydberg macrodimer mode.
	For a pair of sites $(j,k)$ separated by
	\begin{equation}
	    R_{jk}=|x_j-x_k| ,
	\end{equation}
	we denote by $d_{jk,\alpha}^\dagger$ the creation operator of a macrodimer in vibrational/electronic branch $\alpha$, with centre of mass near $(x_j+x_k)/2$ and relative coordinate localized near the molecular bond length. 
	The operator $d_{jk,\alpha}^\dagger$ does not create an ordinary atom on an auxiliary site. 
	It creates a two-atom Rydberg molecular excitation associated with the pair $(j,k)$. 
	Since it contains two fermions, it is bosonic to a good approximation in the low-excitation manifold.

	The target dissipator we seek is
	\begin{equation}
	    \dot\rho = \sum_{j,r} \Gamma_r\,{\mathcal D}[c_jc_{j+r}](\rho) ,
	    \label{eq:target_dissipator}
	\end{equation}
	where
	${\mathcal D}[L](\rho) = L\rho L^\dagger - \frac{1}{2}\left\{L^\dagger L,\rho\right\}$.
	For $r=1$ this is nearest-neighbour pair loss, and for $r=2$ it is next-nearest-neighbour pair loss. 
	In one dimension, the fermionic signs associated with $c_jc_{j+r}$ are fixed by the chosen site ordering. 
	For a dissipator ${\mathcal D}[c_jc_{j+r}]$, an overall sign of the jump operator is immaterial.

	\subsection{Macrodimer association and distance selectivity}
	At the effective level, the association process is described by
	\begin{equation}
	    H_{\rm assoc}^{(\alpha)}
	    =
	    \hbar \sum_{j<k}
	    \left[
	    g_\alpha(R_{jk}) d_{jk,\alpha}^\dagger c_j c_k + {\rm H.c.}
	    \right]
	    + \hbar \sum_{j<k} \delta_\alpha(R_{jk}) d_{jk,\alpha}^\dagger d_{jk,\alpha}.
	    \label{eq:macrodimer_effective_association}
	\end{equation}
	Here $g_\alpha(R_{jk})$ is the effective two-atom association matrix element, including the optical Rabi frequencies and the Franck--Condon overlap between the pinned atom-pair wavefunction and the selected macrodimer wavefunction.
	The detuning
	\begin{equation}
	    \delta_\alpha(R)
	    = 2\varepsilon_\alpha + \frac{U_{u_\alpha}(R)}{\hbar} + \delta_{\rm AC}^{(\alpha)}
	    \label{eq:macrodimer_distance_detuning}
	\end{equation}
	is the two-photon detuning from the molecular resonance.
	In this expression, $\varepsilon_\alpha$ is the single-atom detuning of tone $\alpha$ from the bare Rydberg transition, $U_{u_\alpha}(R)$ is the Rydberg-pair molecular potential for the selected branch $u_\alpha$, and $\delta_{\rm AC}^{(\alpha)}$ denotes light shifts and other experimentally calibrated offsets.
	The sign convention in Eq.~\eqref{eq:macrodimer_distance_detuning} is not essential; what matters is that the resonance condition is
	\begin{equation}
	    \delta_\alpha(R)=0 .
	    \label{eq:macrodimer_resonance_condition}
	\end{equation}

	Equation~\eqref{eq:macrodimer_resonance_condition} is the origin of distance selectivity.
	Since $U_{u_\alpha}(R)$ depends strongly on the interatomic separation, changing the tone frequency changes which separation is resonant. 
	In a uniform one-dimensional array with neighbouring tweezer spacing $s$, a tone satisfying
	\begin{equation}
	    2\varepsilon_1+\frac{U_{u_1}(s)}{\hbar}\simeq0
	    \label{eq:NN_macrodimer_resonance}
	\end{equation}
	addresses nearest-neighbour pairs $(j,j+1)$, whereas a second tone satisfying
	\begin{equation}
	    2\varepsilon_2+\frac{U_{u_2}(2s)}{\hbar}\simeq0
	    \label{eq:NNN_macrodimer_resonance}
	\end{equation}
	addresses next-nearest-neighbour pairs $(j,j+2)$. 
	Thus the frequency of the association tone selects a class of pair separations rather than a unique bond. 
	A global tone resonant at $R=s$ couples to all illuminated nearest-neighbour bonds, while a global tone resonant at $R=2s$ couples to all illuminated next-nearest-neighbour bonds.

	The association strength is appreciable only when the pinned atom pair has non-negligible overlap with the selected macrodimer wavefunction. 
	We may write schematically
	\begin{equation}
	    g_\alpha(R) \propto \Omega_\alpha^{\rm eff} \eta_{u_\alpha}(R),
	    \label{eq:FC_pair_coupling}
	\end{equation}
	where $\Omega_\alpha^{\rm eff}$ contains the relevant optical coupling strengths and $\eta_{u_\alpha}(R)$ is a Franck--Condon factor. 
	The sharp dependence of both $\delta_\alpha(R)$ and $\eta_{u_\alpha}(R)$ on $R$ underlies the experimentally observed distance selectivity of macrodimer excitation and dressing \cite{Hollerith2019,Hollerith2022}. 
	In practice, the usable selectivity is limited by the molecular linewidth, the association Rabi frequency, the motional width of the pinned atoms, and possible off-resonant coupling to nearby vibrational resonances.

	For a spatially uniform drive, Eq.~\eqref{eq:macrodimer_effective_association} should be understood as a sum over local macrodimer modes. 
	A nearest-neighbour tone and a next-nearest-neighbour tone generate
	\begin{equation}
	    H_{\rm assoc}^{(1)}
	    = \hbar \sum_j \left[ g_1 d_{j,j+1}^{(1)\dagger}c_jc_{j+1} + g_2 d_{j,j+2}^{(2)\dagger} c_j c_{j+2} +
	    {\rm H.c.}
	    \right]
	    + \hbar \sum_j \delta_1 d_{j,j+1}^{(1)\dagger}d_{j,j+1}^{(1)}  + \sum_j \delta_2 d_{j,j+2}^{(2)\dagger}d_{j,j+2}^{(2)}
	    .
	    \label{eq:NN_effective_macrodimer_association}
	\end{equation}
	The internal molecular resonance is the same on every bond, but $d_{j,j+1}^{(1)}$ and $d_{j+1,j+2}^{(1)}$ are distinct local modes with different centre-of-mass positions.

	\subsection{Converting the macrodimer into a lossy auxiliary mode}

	To realize irreversible two-body loss, the macrodimer mode must be made short-lived. 
	We write the molecular decay as
	 $   \kappa_\alpha {\mathcal D}[d_{jk,\alpha}](\rho) $.
	The linewidth $\kappa_\alpha$ may contain intrinsic Rydberg decay, motional escape from the trap, photoionization, or an engineered dump channel. 
	A useful controlled mechanism is to apply an additional dump tone coupling the macrodimer state $|M\rangle$ to a rapidly decaying state $|B\rangle$,
	\begin{equation}
	    |M_{jk}^{(u_\alpha)}\rangle
	    \xrightarrow{\Omega_{D,\alpha}} |B_{jk}\rangle \xrightarrow{\gamma_B}{\mathrm {loss}}.
	\end{equation}
	The corresponding Hamiltonian is
	\begin{equation}
	    H_D/\hbar
	    = \Delta_B |B\rangle\langle B| + \frac{\Omega_{D,\alpha}}{2} \left( |B\rangle\langle M| + |M\rangle\langle B|
	    \right),
	\end{equation}
	with a dissipator $\gamma_B{\mathcal D}[|{\rm loss}\rangle\langle B|](\rho)$. 
	Eliminating the rapidly decaying state $|B\rangle$ gives an induced macrodimer linewidth
	\begin{equation}
	    \kappa_{\alpha,{\rm dump}} \simeq \frac{ \gamma_B |\Omega_{D,\alpha}|^2}{ 4\Delta_B^2+\gamma_B^2 }.
	    \label{eq:kappa_dump}
	\end{equation}
	Thus the linewidth is tunable by the dump-tone intensity and detuning. 
	Controlled laser-induced Rydberg dissipation by coupling a Rydberg level to a short-lived low-lying excited state has recently been demonstrated for atomic Rydberg states~\cite{Begoc2025}; applying this idea to a selected macrodimer manifold is a natural reservoir-engineering extension.

	For a single pair channel, the auxiliary master equation is
	\begin{align}
	    \dot\rho = -\frac{i}{\hbar} [H_{jk,{\rm eff}}^{(\alpha)},\rho] + \kappa_\alpha {\mathcal D}[d_{jk,\alpha}](\rho) .
	    \label{eq:auxiliary_master}
	\end{align}
	If
	\begin{equation}
	    \kappa_\alpha,\ |\delta_{\alpha,{\rm eff}}| \gg |g_\alpha|,
	\end{equation}
	the macrodimer is only virtually populated and can also be adiabatically eliminated. 
	The Heisenberg equation gives
	\begin{equation}
	    d_{jk,\alpha} \simeq - \frac{i g_\alpha}{ i\delta_{\alpha,{\rm eff}}+\kappa_\alpha/2}c_jc_k ,
	\end{equation}
	so that the effective jump operator acting on the ground-state fermions is
	\begin{equation}
	    L_{jk}^{(\alpha)} = \sqrt{\Gamma_\alpha(R_{jk})}\, c_jc_k ,
	\end{equation}
	with
	\begin{equation}
	    \Gamma_\alpha(R) = \frac{\kappa_\alpha |g_\alpha(R)|^2}{\delta_{\alpha,{\rm eff}}(R)^2 + \kappa_\alpha^2/4}.
	    \label{eq:Gamma_pair}
	\end{equation}
	On resonance, $\delta_{\alpha,{\rm eff}}=0$, this reduces to
	\begin{equation}
	    \Gamma_\alpha = \frac{4|g_\alpha|^2}{\kappa_\alpha}.
	\end{equation}
	Thus increasing $\kappa_\alpha$ indefinitely is not optimal: in the strongly overdamped limit the effective pair loss is suppressed by a quantum-Zeno mechanism.

	\subsection{Collective pair loss from two macrodimers coupled to one lossy mode}

	We now describe a structured reservoir that realizes a coherent superposition of two pair-annihilation operators. 
	Let
	\begin{equation}
	    A_1=c_{j+1}c_{j+2}, \qquad A_2=c_{j}c_{j+3}.
	\end{equation}
	The two pairs have the same centre of mass in a uniform chain but different relative separations, $s$ and $3s$. 
	Suppose that $A_1$ couples to a macrodimer $d_1$ and $A_2$ couples to a second macrodimer $d_2$:
	\begin{equation}
	    H_{\rm pair}/\hbar
	    = \sum_{\mu=1,2} \left[ \delta_\mu d_\mu^\dagger d_\mu + g_\mu d_\mu^\dagger A_\mu + g_\mu^* A_\mu^\dagger d_\mu \right].
	    \label{eq:two_dimer_pair_hamiltonian}
	\end{equation}
	The two macrodimers are then coupled to a common short-lived mode $b$,
	\begin{equation}
	    H_b/\hbar = \delta_b b^\dagger b + \sum_{\mu=1,2} \left[ J_\mu b^\dagger d_\mu + J_\mu^* d_\mu^\dagger b \right],
	    \label{eq:common_loss_hamiltonian}
	\end{equation}
	with $\kappa_b{\mathcal D}[b](\rho)$.
	The mode $b$ may represent a rapidly decaying electronic molecular state, an ionization continuum, or a deliberately engineered dump channel. 
	The essential requirement is that decay from $b$ does not reveal whether $d_1$ or $d_2$ was populated.

	In the limit
	\begin{equation}
	    \kappa_b\gg |J_1|,|J_2|,|\delta_b|,
	\end{equation}
	$b$ can be eliminated, giving a collective decay channel for the two macrodimers,
	\begin{equation}
	    \dot\rho_d = \frac{4}{\kappa_b} {\mathcal D}\!\left[ J_1d_1+J_2d_2 \right](\rho_d) ,
	    \label{eq:collective_dimer_loss}
	\end{equation}
	up to small coherent Lamb shifts. If the macrodimers themselves are only virtually populated, they can also be eliminated. The resulting jump operator acting directly on the target fermions is
	\begin{equation}
	    L_j = \sqrt{\frac{4}{\kappa_b}} \left[ J_1 \chi_1 g_1\, c_jc_{j+1} + J_2 \chi_2 g_2\, c_{j-1}c_{j+2} \right],
	    \label{eq:collective_jump_general}
	\end{equation}
	where
	\begin{equation}
	    \chi_\mu = \frac{1}{\delta_\mu-i\gamma_\mu/2}
	\end{equation}
	is the susceptibility of macrodimer $\mu$, including any residual linewidth $\gamma_\mu$. By tuning the intensities and optical phases of the association and dump tones, one may choose $J_1\chi_1g_1 = J_2\chi_2g_2$,
	which gives desired dissipator
	\begin{equation}
	    L_j = \sqrt{\Gamma_{\mathrm col}} \left( c_jc_{j+1} + e^{i\phi} c_{j-1}c_{j+2} \right).
	    \label{eq:desired_collective_jump}
	\end{equation}

	\subsection{Projected single-particle pump}

	We now describe a possible implementation of a local single-particle pump and its nearest-neighbour-blockaded extension.
	A microscopic realization can be obtained by coupling target site $i$ to an auxiliary source mode $s_i$.
	The source mode is assumed to be continuously replenished from a reservoir and rapidly refilled.
	A Raman-assisted tunnelling process gives
	\begin{equation}
	    H_{\mathrm {inj}} = \hbar J_i \left( c_i^\dagger s_i+s_i^\dagger c_i \right),
	    \label{eq:source_target_hamiltonian}
	\end{equation}
	where $s_i$ annihilates a fermion in the source mode and $J_i$ is the source-to-target coupling matrix element.
	The source is maintained close to an occupied, Markovian state.
	Eliminating the source in the limit where its refilled rate $\kappa_s$ is the fastest scale gives the effective gain process
	\begin{equation}
	    \dot\rho_{\rm target} = \gamma_i{\mathcal D}[c_i^\dagger](\rho_{\rm target}), \qquad \gamma_i\simeq \frac{4|J_i|^2}{\kappa_s}
	    \label{eq:source_elimination_rate}
	\end{equation}
	on resonance.
	More generally, if the source-to-target injection is detuned by $\delta_i$, the pump rate has the Lorentzian form
	\begin{equation}
    	\gamma_i(\delta_i) = \frac{\kappa_s |J_i|^2} {\delta_i^2+\kappa_s^2/4}.
    	\label{eq:detuned_pump_rate}
	\end{equation}

	Experimentally, the source mode can be a local auxiliary tweezer, buffer trap,
	or reservoir-coupled tweezer that supplies atoms to the target register.
	Such reservoir-based loading architectures have been demonstrated for tweezer arrays and provide the atom-supply ingredient needed for a local pump~\cite{PhysRevResearch.5.L032009}.  

	To block the pump when a neighbouring target site is occupied, the energy for adding an atom to site $i$ must depend on the occupations of nearby sites.
	This can be achieved by off-resonantly dressing the target atoms with a Rydberg state.
	Let $|g\rangle$ be the target electronic ground state and $|r\rangle$ a Rydberg state.
	A dressing tone with Rabi frequency $\Omega_{\rm d}$ and detuning $\Delta_{\rm d}$ couples
	\begin{equation}
	    |g\rangle \leftrightarrow |r\rangle, \qquad |\Delta_{\rm d}|\gg |\Omega_{\rm d}|.
	\end{equation}
	The dressed ground state contains a small Rydberg admixture,
	\begin{equation}
	    |\tilde g\rangle \simeq |g\rangle + \frac{\Omega_{\rm d}}{2\Delta_{\rm d}}|r\rangle.
	\end{equation}
	Because two Rydberg atoms interact, two occupied dressed ground-state sites acquire an effective interaction.
	At the target-mode level this is described by
	\begin{equation}
	    H_{\rm dress} = \sum_{m<n} V_{\rm dress}(R_{mn})\,n_m n_n ,
	    \label{eq:rydberg_dressed_interaction}
	\end{equation}
	where $R_{mn}$ is the distance between target sites $m$ and $n$.
	The function $V_{\rm dress}(R)$ may be a conventional soft-core Rydberg-dressed interaction or, using off-resonant coupling to Rydberg molecular potentials, a
	distance-selective interaction peaked near a chosen separation~\cite{PhysRevA.82.033412,Hollerith2022}.

	The addition energy of site $i$ is therefore shifted by the occupations of the other sites:
	\begin{equation}
	    E_i^{\mathrm{add}} = E_i^0 + \sum_{m\neq i} V_{\mathrm{dress}}(R_{im})\,n_m .
	    \label{eq:addition_energy_general}
	\end{equation}
	For a three-site system $(i-1,i,i+1)$ with neighbouring target-tweezer spacing $s$, this becomes
	\begin{equation}
	    E_i^{\mathrm{add}} = E_i^0 + V_{\mathrm{dress}}(s) \left( n_{i-1}+n_{i+1} \right),
	    \label{eq:addition_energy_three_site}
	\end{equation}
	up to smaller longer-range shifts.
	The dressing tone must address the target atoms on all three tweezers, because the blockade arises from the interaction between a possible atom on the middle site and atoms already occupying the neighbouring sites.

	The source-to-target injection tone is chosen to be resonant with the addition energy of the middle site when both neighbours are empty,
	\begin{equation}
	    \hbar\omega_{\rm inj}=E_i^0 .
	\end{equation}
	Using Eq.~\eqref{eq:addition_energy_three_site}, the configuration-dependent injection detuning is
	\begin{equation}
	    \delta_i(n_{i-1},n_{i+1}) = \frac{ V_{\rm dress}(s_{\rm lat})}{\hbar} \left(n_{i-1}+n_{i+1}\right).
	    \label{eq:configuration_dependent_detuning}
	\end{equation}
	Substituting this into Eq.~\eqref{eq:detuned_pump_rate} gives
	\begin{equation}
	    \gamma_i(n_{i-1},n_{i+1}) = \frac{\kappa_s |J_i|^2} { \left[ V_{\rm dress}(s) (n_{i-1}+n_{i+1})/\hbar \right]^2 + \kappa_s^2/4}.
	    \label{eq:configuration_dependent_rate}
	\end{equation}
	When both neighbouring sites are empty,
	\begin{equation}
	    \gamma_i(0,0) = \frac{4|J_i|^2}{\kappa_s} \equiv \gamma_i.
	\end{equation}
	If one neighbour is occupied, the pump is detuned by $V_{\mathrm{dress}}(s)/\hbar$.
	If both neighbours are occupied, it is detuned by $2V_{\mathrm{dress}}(s)/\hbar$.
	In the blockade regime
	\begin{equation}
	    |V_{\rm dress}(s_{\rm lat})| \gg \hbar |J_i|,\ \hbar\kappa_s, \label{eq:blockade_condition_pump}
	\end{equation}
	the off-resonant rates are strongly suppressed:
	\begin{equation}
	    \gamma_i(1,0),\ \gamma_i(0,1),\ \gamma_i(1,1) \ll \gamma_i(0,0).
	\end{equation}
	Thus the effective jump operator becomes
	\begin{equation}
	    L_i^{\mathrm{blockaded}} = \sqrt{\gamma_i}\,c_i^\dagger (1-n_{i-1})(1-n_{i+1}) .
	    \label{eq:blockaded_pump_jump}
	\end{equation}
	The corresponding master equation is
	\begin{equation}
	    \dot\rho = \gamma_i\ {\mathcal D} \!\left[ c_i^\dagger (1-n_{i-1})(1-n_{i+1})\right](\rho).
	    \label{eq:blockaded_pump_master}
	\end{equation}
	Equation~\eqref{eq:blockaded_pump_master} fills the middle target tweezer only when the middle site is empty and both neighbouring target sites are empty.  The emptiness of the middle site is enforced automatically by fermionic Pauli blocking, while the emptiness of the neighbouring sites is enforced by the
	Rydberg-dressed interaction shift.